\documentclass[conference]{IEEEtran}
\usepackage{graphicx}
\ifCLASSINFOpdf
\else
\fi
\begin{document}
%
\title{Low Precision Constant Parameter CNN on FPGA}

\author{
\IEEEauthorblockN{Hah, Thiam Khean; Liew, Yeong Tat; Jason Ong}
\IEEEauthorblockA{Intel PSG}
\IEEEauthorblockA{\{thiam.khean.hah; yeong.tat.liew; jason.gee.hock.ong\}@intel.com}
}


%


\maketitle

\begin{abstract}
We report FPGA implementation results of low precision CNN convolution layers optimized for sparse and constant parameters. We describe techniques that amortizes the cost of common factor multiplication and automatically leverage dense hand tuned LUT structures. We apply this method to corner case residual blocks of Resnet on a sparse Resnet50 model to assess achievable utilization and frequency and demonstrate an effective performance of 131 and 23 TOP/chip for the corner case blocks. The projected performance on a multichip persistent implementation of all Resnet50 convolution layers is 10k im/s/chip at batch size 2. This is 1.37x higher than V100 GPU upper bound at the same batch size after normalizing for sparsity.
\end{abstract}


%
\IEEEpeerreviewmaketitle

\section{Introduction and Related Works}
In recent years, Convolutional Neural Networks(CNN) have demonstrated great efficacy on computer vision tasks such as classification\cite{resnet}, localization\cite{yolo}, and SRGAN\cite{srgan}. Together with Recurrent Neural Networks(RNN), it has motivated the development of custom silicon for Deep Learning(DL). For example, GPU Tensor Core\cite{volta}, TPU\cite{tpu} and Graphcore\cite{graphcore}. 

There has also been work to optimize DL on programmable logic. Notably, Song Han et al.\cite{fpga_codesign} proposed software-hardware co-design. While silicon implementations must customize for a range of DL applications, an FPGA can customize to a single DL application. This enables application specific customization of precision, sparsity and network structure. However, this is not the limit of FPGA customization. FPGAs can be further customized to a specific instance of a DL application by implementing post training parameters as constants. We call this a Compiled CNN or RNN. 

This paper reports the results of an early investigation into Compiled CNNs for sparse low precision models. We also automate the process of converting quantized caffemodels into an equivalent hardware design built using efficient hand tuned Intel FPGA WYSIWYGS. In this early work, we implementation only the predicted corner case blocks of Resnet50.

\section{Compiled CNN Implementation}

\subsection{Resnet50 Model Sparsity and Precision}
An advantage of Compiled CNNs is its ability to exploit fine grained parameter sparsity without overhead. Multiply-Accumulates (MAC) associated with constant zeros are simply dropped. AMC\cite{amc} showed 80\% sparse Resnet50 with no accuracy loss. We use an 80\% sparse model from Movidius\cite{movidius_sparse} to us as a proxy. We received a pre-quantized model obtained using a modified version of TRN\cite{trn} as our starting point. In version of TRN, each output channel has one independent scaling factor and 6 residual terms (equivalent to INT7) are used to obtain an accuracy loss of just 0.22\% vs FP32.

\subsection{Selection of Resnet Layers for Implementation}
Table \ref{design_param} shows that the key design parameters of conv3\_x and conv4\_x are bounded by those of conv2\_x and conv5\_x. The design corners are therefore represented by conv2\_x and conv5\_x. We focus our efforts on these 2 layers to as a way to quickly assess the potential of Low Precision Compiled CNNs.

\begin{table}[ht]
\caption{Key Design Parameters}
\label{design_param}
\begin{center}
\begin{tabular}{|c|c|c|c|c|}
\hline
Layer & conv2\_x & conv3\_x & conv4\_x & conv5\_x \\
\hline
Channel Count & 64/256 & 128/512 & 256/1024 & 512/2048 \\
\hline
Channel Height/Width & 56x56 & 28x28 & 14x14 & 7x7  \\
\hline
Parameter Count(k) & 69 & 279 & 1114 & 4456 \\
\hline
Total MACs(M) & 218 & 218 & 218 & 218 \\
\hline
Total MAC/Parameter & 3136 & 784 & 196 & 49 \\
\hline
\end{tabular}
\end{center}
\end{table}

\subsection{Top Level}

\graphicspath{ {./pics/} }
\begin{figure}[ht]
\centering
\includegraphics[scale=0.45]{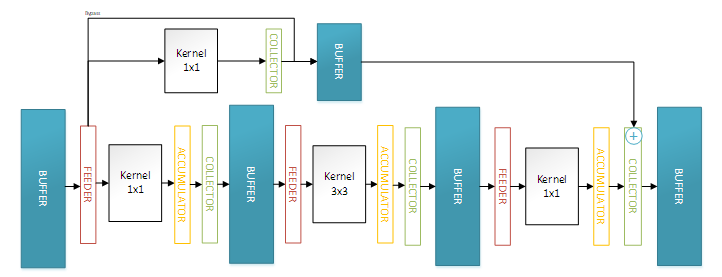}
\caption{Resnet Residual Block Implementation}
\label{residual_block}
\end{figure}

The basic unit of design is the Resnet Residual Block(Fig \ref{residual_block}). We require that this fits on a single chip to keep residual shortcut data on chip. The top level is divided into 2 modules. The Kernel implements the CNN Multiply-Accumulates(MAC). Everything else is part of the Non Kernel(NK). The benefits of constant parameter optimization is found mostly in the Kernel and we focus our efforts there. The NK is, at present, modestly optimized. Where needed we fill the GX280 with several slightly modified Residual Blocks to evaluate resource use, frequency and routability at high utilization.

Compiling post training parameters into FPGAs yields a persistent network. Therefore, we expect a multichip implementation with suitable interchip links, like, Ethernet or PCIE. To minimize the number of chips needed, we use bit serial math in the Kernel. While bit serial operations are slower they are also smaller and consequently more numerous. In theory, bit serial math does not reduce performance per Logic Element(LE). In practice, it is lower because typical implementations are unable to use the hard adders and carry chains built into modern FPGAs. We will present a solution to overcome inefficiency.

\subsection{Non Kernel Design}

The NK module includes everything that is not part of a single convolutional step. This includes buffers for intermediate feature maps, data movement between Kernels and other operations such as bias add, activation functions, normalization and rounding. It also performs the partial result accumulation across convolutional steps for filter sizes greater than 1x1.

We intend to automatically generate the NK RTL in the future but it is currently hand coded. We have also set aside resources for implementing the interchip Ethernet/PCIE channels but have not yet done so.

\subsubsection{Buffers}
The Buffers are constructed using FPGA block RAMs as either (a) streaming FIFOs or (b) double buffers. Buffer type is selected based on layer dataflow, FPGA resource allocation and time division multiplexing (TDM) of the kernel operation. In all cases, interchip buffers are double buffers.

The residual shortcut shown on top of the Fig \ref{residual_block} may have an additional buffer and 1x1 kernel if needed. The buffers at the input and output boundary of a Residual Block is shared with the prior and next Residual Block respectively if they are on the same chip. Otherwise, the buffers also serve as the chip level input and output buffer. 

\subsubsection{Feeder}
The feeder fetches parallel data stored in the Buffer and serializes it to feed the bit serial Kernel as well as the residual shortcuts. Deserialization is not needed because the bit serial Kernel generates a parallel output.

\subsubsection{Accumulator}
The Accumulator block adds the partial sums across multiple convolutional steps and is not required for layers with 1x1 filters. When the final sum is ready, it streams the sum to the Collector block. 

\subsubsection{Collector}
The collector performs miscellaneous operations such as bias addition, scaling (normalization) and ReLU. The last collector within each Residual block also adds the shortcut data. The activations are also saturated and rounded to 8 bits here. The design uses as many bit as necessary at other stages to avoid rounding/saturation elsewhere and we make use of constant parameter and ReLU properties to minimize the number of bits needed at every stage. DSP blocks used for scaling are shared by multiple Output Feature Maps(OFM). This is possible because bit serial math requires multiple clocks per operation while DSPs require only one.

\subsection{Kernel Design Compilation and Flow}
\begin{figure}[ht]
\centering
\includegraphics[scale=0.6]{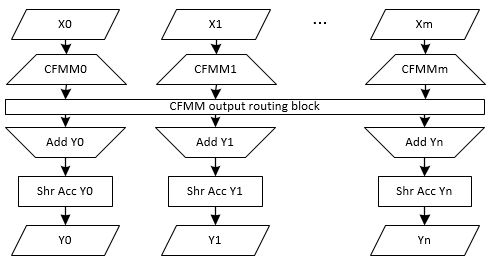}
\caption{Kernel implementation}
\label{kernel_abstract_diagram}
\end{figure}

We developed a tool to automatically convert the constant parameters of a single convolution layer stored in a caffemodel into the Kernel RTL optimized for Intel Stratix 10. The NK consumes the RTL as a blackbox and is unaffected by the changes in parameter values and retraining. The Kernel design shown in Fig \ref{kernel_abstract_diagram} is composed of the module inputs and outputs (Xm and Yn), the Common Factor Mass Multiplication (CFMM) blocks, bit serial adder tree (Add Yn) and shift right accumulator (Shr Acc Yn).

\graphicspath{ {./pics/} }

\subsubsection{Common Factor Mass Multiplication (CFMM) Blocks}
\begin{figure}[ht]
\centering
\includegraphics[scale=0.38]{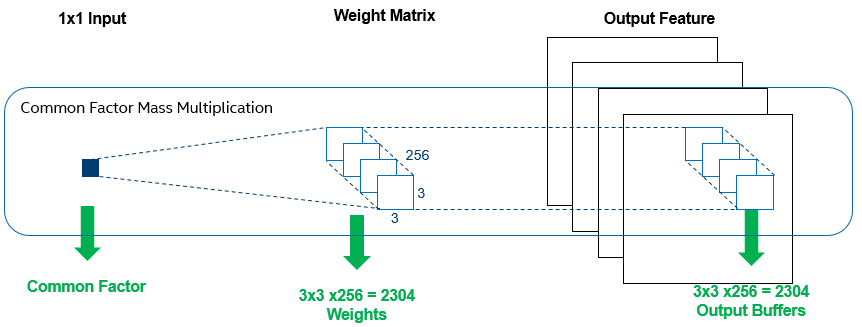}
\caption{CFMM Common Factor}
\label{cfmm_common_factor}
\end{figure}

We reduce the cost of multiplication by amortizing it across multiple operations sharing the same Common Factor(CF). To enable this, we refactor the computation to take a set of inputs from an input feature map (IFM) and perform all computations for it in a single pass. This turns the inputs values into a CF. Fig \ref{cfmm_common_factor} shows an example, where a single input value acts as the CF for 2304 multiplications (3x3 filter with 256 OFM). 

\graphicspath{ {./pics/} }

If the multiplier values are well distributed and numerous relative to the number of unique products, then every product is likely needed and the optimal CFMM design is trivial. The will show that the number of unique multiplications required for an INT7 parameter is 32 for Compiled CNNs. This is small relative to the typical number of CF multiplications in CNNs even at 80\% sparsity. We optimize the design for this case and allow the vendor tools to remove unused output products during synthesis, should they exist.

To minimize the number of unique products to compute, we move the sign bit of the parameter into the adder tree. This equivalence classes positive and negative values and reduces the number of unique INT7 products to 64. Also, even products can be produced with a shift left of an odd product and constant shift lefts are free on FPGAs (costs 1 flop for bit serial math). Thus a INT7 CFMM block only has 32 unique products and multiplication by 0 and 1 is free. Now note that, (a) when generating all products each \textit{incremental} product can be generated using one add/sub and (b) the cost of a bit serial adder is about 1 ALM. Therefore, the first order cost of a CFMM block only about 30 ALMs plus flops. This renders the \textit{ALM} cost of multiplication trivial. However, each product must still be routed to the an adder tree. This makes FPGA routablity a fundamental limiter on the efficiency of CFMM based multiplication.

\subsubsection{Adder Tree and Shift Right Accumulator}
\begin{figure}[ht]
\centering
\includegraphics[scale=0.3]{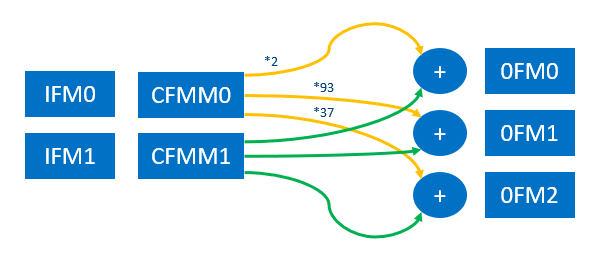}
\caption{OFM Adder Trees add selected constant products from CFMM Blocks}
\label{adder_tree}
\end{figure}

Each CFMM block computes a product for all OFMs for one IFM. However, each OFM output is a sum of the products from multiple IFMs. Therefore, the design requires one CFMM per IFM and one adder tree for each OFM to sum the products from all CFMM blocks. A simple example is shown in Fig \ref{adder_tree}.

\graphicspath{ {./pics/} }

\begin{figure}[ht]
\begin{center}
\includegraphics[scale=0.34]{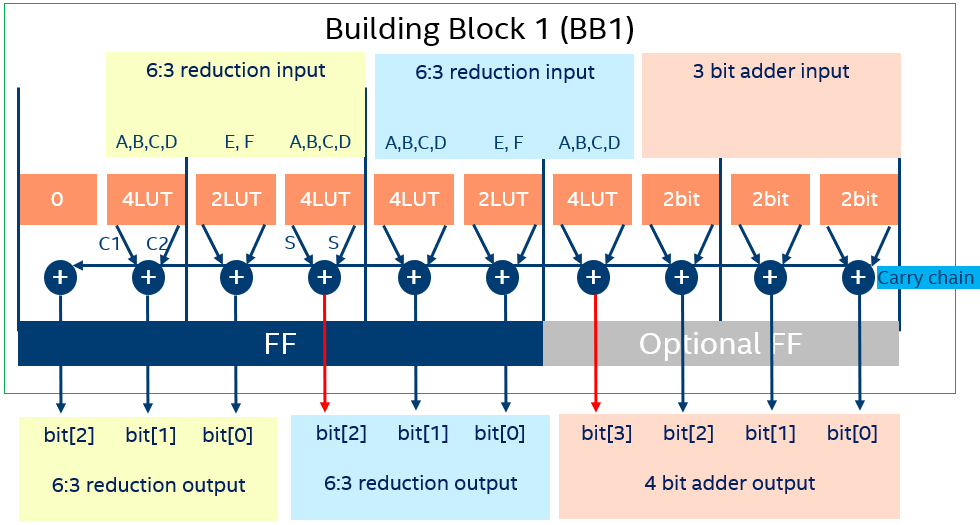}
\caption{6:3 Reduction with Carry Hiding.}
\label{6_3_reduction}
\end{center}
\scriptsize{This contains 2 6:3 reduction blocks(blue and yellow) and a 3 bit adder(red). Each 6:3 reduction block adds 6 bits(A to F). E+F are simple adds. The addition of A,B,C,D results in 2 Carry bits (encodes the value 4) and 1 Sum bit. The half ALM to the left of E+F computes the 2 Carry bits and adds them. The half ALM to the right computes 0.5*(Sum bit) and adds \textit{2} of them to yield the Sum bit. The 3 bit adder adds the outputs from the 2 6:3 reduction blocks into a 4 bit sum. Subsequent stages in the adder tree uses parallel adders. The red lines indicate hidden carries. Here, a single half ALM performs 2 orthogonal functions. It (a) computes the Sum bit of A+B+C+D and transmits it on the ALM Carry Out and (b) transmits the carry in from the \textit{previous} adder on the ALM Sum Out. This hides the carry of a previous adder in the shadow of another computation, making it free. Quartus implements this verbatim and reports timing loops on false paths.}
\end{figure}

With cheap CFMM multipliers, adder trees become the main consumer of ALMs. However, bit serial math is poorly supported by current FPGA architecture and tools. We find that vendor tools implement 3:2 reduction and bit serial math at half the efficiency of parallel adds. To resolve this, we introduce (a) a hand tuned WYSIWYG design for Intel Stratix 10 with (b) carry hiding. It performs a 6:3 reduction in one ALM stage and asymptotically uses only 3 ALMs. This is double the efficiency of vendor tools and brings the efficiency of bit serial adders back inline with parallel adders. Fig \ref{6_3_reduction} describes a 5 ALM variant of this structure for adding 12 bits. Several variants of this design were built as hand coded WYSIWYG modules which are automatically instantiated by tools converting caffemodels to RTL. This leverages the efficiency of hand tuned designs while hiding its complexity.

The largest variants uses 10 ALMs and add up to 27 bits. While denser variants uses of fewer ALMs it may negatively impact routability and frequency. To determine the optimal variant, we sweep the variants with parameters such as pipelining and accumulator reset strategy. We find that the variant in Fig \ref{6_3_reduction} with one pipeline stage for every 2nd adder stage best meet our frequency and routability requirements.

\graphicspath{ {./pics/} }

Also, the adder tree adds a single bit in a bit serial value every clock. To get the final sum, a shift right accumulator(SRA) is added to the end of the adder tree. The SRA simply shifts right the sum of bit N and accumulates it with the sum for bit N+1. For efficiency, the adder trees perform 1's complement math. A constant modifier in the NK module's bias adder converts the final sum into 2's complement for free.

\subsubsection{Multi Instance Kernels and Folding}

Compiled CNNs requires that every parameter be hardened into the FPGA. Table \ref{design_param} shows that the amount of compute per parameter differ by up to 64x between conv2\_x and conv5\_x. To maintain the same throughput at every layer we may need to fold (TDM) the kernels or use multiple instances as appropriate.  

\begin{figure}[ht]
\centering
\includegraphics[scale=0.6]{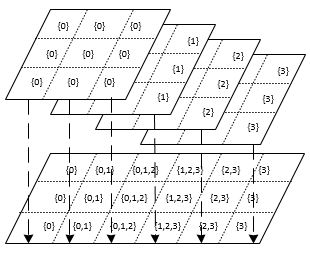}
\caption{Multi-instance Kernel with 4 3x3 filters summed into a 3x6 output}
\label{multi_inst_kernel}
\end{figure}

Kernel folding is implemented using muxes. As a result, the almost free CFMM multiplication now requires a mux each. In multi instance kernels, each instance corresponds to one convolutional step. For efficiency, the multi instance Kernels directly sum the partial products across multiple convolutional steps. Fig \ref{multi_inst_kernel} illustrates this for a 4 instance kernel with a 3x3 filter summed into a 3x6 output slice. 

\graphicspath{ {./pics/} }

\section{Implementation Targets and Results}
\subsubsection{Residual Block on GX280}
We implement conv2\_2 and conv5\_2 using the techniques described above. As noted, we require that Residual Blocks be fully contained within one chip. Initial experiments show that conv5\_2 must be folded by 4x to fit on GX280. Also, 8 instances of conv2\_2 kernels at 2x the frequency of the conv5\_2 layer is required to match their inference throughput. This is implemented as 2 conv2\_2 Kernel modules with 4 instances arranged as in Fig \ref{multi_inst_kernel}. 

The design targets and implementation results are summarized in Table \ref{target_and_result}. To simulate high chip utilization we duplicated the conv2\_2 Kernels 5 times. This is not needed for conv5\_2. However, conv5\_2 Kernels contains duplicates of each CFMM block to alleviate routing congestion. The effective TOP/Chip reports the number of effective TOPs which includes the benefits of sparsity.

\begin{table}[ht]
\caption{Design Targets and Implementation Results}
\label{target_and_result}
\begin{center}
\begin{tabular}{|c|c|c|c|c|}
  \hline
    & \multicolumn{2}{|c|}{Target} & \multicolumn{2}{|c|}{Actual} \\
  \hline
    Layer & conv2 & conv5 & conv2 & conv5 \\
  \hline
    Instances/Kernel & 4 & 1 & 4 & 1 \\
  \hline
    Folding & 1 & 4 & 1 & 4 \\
  \hline
    Frequency(MHz) & 400 & 200 & 353 & 156 \\
  \hline
    Chip Utilization & \- & \- & 76\% & 67\% \\
  \hline
    ALM/Kernel(k) & \- & \- & 127 & 620 \\
  \hline
    DSP/Kernel & \- & \- & 96 & 256 \\
  \hline
    M20K/Kernel & \- & \- & 1852 & 1100 \\
  \hline
    CFMM Dupe & 1 & 1 & 1 & 2 or 4 \\
  \hline
    Effective MOPs/ALM & 80 & 16 & 70 & 12 \\
  \hline
    GX280 Effective TOPs/Chip & 74 & 15 & 66 & 12 \\
  \hline
    GX550 Effective TOPs/Chip* & 149 & 30 & 131 & 23 \\
  \hline
\end{tabular}
\end{center}
  \scriptsize{* : Projection}
\end{table}

\subsubsection{Projections and Comparisons}

TOP/chip of GX280 is an inaccurate measure of the fundamental FPGA capability. Compiled CNNs are DSP light and use <20\% of the DSPs on DSP heavy GX280. At the same performance density, the DSP light GX550 would yield 131 and 23 TOP/chip for conv2\_2 and conv5\_2 respectively. 

Comparing the largest monolithic FPGA against the largest monolithic GPU, we see that the conv2\_2 performance of 131 TOP/chip compares favorably with NVidia V100\cite{volta} with a \textit{peak} performance of 125 TOP/chip . Poor conv5\_2 performance is the result of Kernel folding. An FPGA with sufficiently numerous ALMs would avoid the need for folding and result in higher TOP/chip which would likely be closer to those seen on conv2\_2. Additionally, it should be noted that the use of Compiled CNNs is not all or nothing. A CNN partitioned into multiple chips can be implemented partly as Compiled CNN and partly through other means. Where practical, a hybrid solution may yield the optimal solution.

\begin{figure}[ht]
\centering
\includegraphics[scale=0.27]{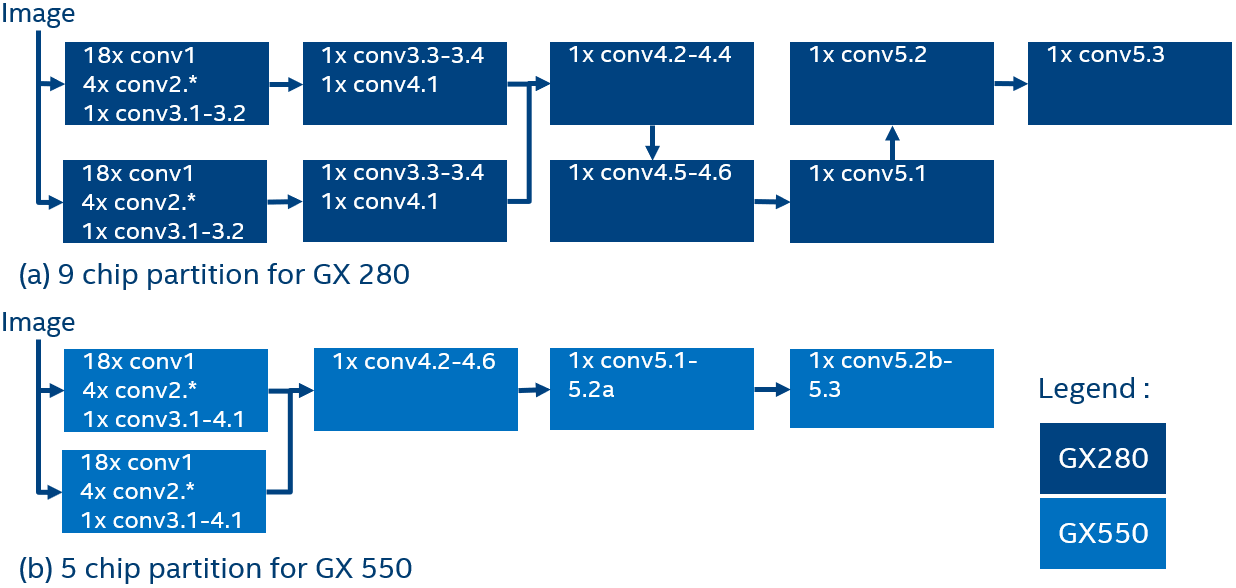}
\caption{Multi Chip Partitioning of Resnet50 (Projection)}
\label{network_partitioning}
\end{figure}

Finally, we use the demonstrated implementation results to estimate the resource requirements for the remaining convolution layers. This was used to create a reasonable multichip partitioning of Resnet50(Fig \ref{network_partitioning}). It is throughput balanced and requires at most ~75Gbps links. At frequencies demonstrated its throughput is ~53061 image/second at batch 2. This corresponds to 5896 and 10612 im/s/chip on GX280 and GX550 respectively. The throughput of a V100 in a DGX-1 system\cite{nvidia_perf} at batch 2 is 1544 im/s/chip. If V100 can extract a 5x efficiency from the 80\% unstructured sparsity in the model, its upper bound performance would be 7720 im/s/chip. Our implementation is 1.37x faster than that bound.

\graphicspath{ {./pics/} }

At submission time, the full system performance is an estimate and has not been validated. Additionally, it does not include the FC layer which we intend to offload to the CPU. However, we feel that the optimistic V100 assumptions more than make up for it and the ability of Compiled CNNs to naturally exploit unstructured sparsity without overhead is a fundamental benefit of this approach. Similarly, the ability to use INT7 with similar accuracy to INT8 and FP32 is a legitimate strength of FPGAs in general.

\section{Future Work}

Future work may include full network implementation plus power and latency measurements which should account for inter chip link power and latency. We note that the low clock frequency is a positive for power. A more complete comparison of our work against sparse persistent GPU and FPGA implementations would also be useful. Finally, additional improvements to Compiled CNNs performance through tools, IP design or FPGA architecture may be explored.

\section{Conclusion}
We proposed the use of Compiled CNNs to improve FPGA efficiency and exploit parameter sparsity during CNN inferencing. We introduced techniques to amortize multiplication cost and automated tools that exploit the efficiency of hand tuned designs. We then demonstrated these techniques on sample corner case residual blocks of Resnet50 and use the results to estimate performance for all Resnet50 convolution layers. The projected performance on GX550 is 10612 im/s/chip which is 1.37x higher than the V100 upper bound at the same batch size after normalizing for sparsity.


\section*{Acknowledgment}
Authors of \cite{movidius_sparse} and \cite{trn} for sparsifying and quantizing the Resnet50 model respectively.



%

\end{document}